# Homogeneous optical cloak constructed with uniform layered structures

Jingjing Zhang[1,†,*], Liu Liu[1,*], Yu Luo[2], Shuang Zhang[3] and Niels Asger Mortensen[1,†]

[1] *DTU Fotonik - Department of Photonics Engineering, Technical University of Denmark, DK-2800 Kongens Lyngby, Denmark*

[2] *The Blackett Laboratory, Department of Physics, Imperial College London, London SW7 2AZ, UK*

[3] *School of Physics and Astronomy, University of Birmingham, Birmingham, UK*

[†] *Authors to whom correspondence should be addressed: jinz@fotonik.dtu.dk; asger@mailaps.org.*

[*] *These two authors contribute equally to this work.*

**The prospect of rendering objects invisible has intrigued researchers for centuries. Transformation optics based invisibility cloak design is now bringing this goal from science fictions to reality and has already been demonstrated experimentally in microwave and optical frequencies. However, the majority of the invisibility cloaks reported so far have a spatially varying refractive index which requires complicated design processes. Besides, the size of the hidden object is usually small relative to that of the cloak device. Here we report the experimental realization of a homogenous invisibility cloak with a uniform silicon grating structure. The design strategy eliminates the need for spatial variation of the material index, and in terms of size it allows for a very large obstacle/cloak ratio. A broadband invisibility behavior has been verified at near-infrared frequencies, opening up new oppotunities for using uniform layered medium to realize invisibility at any frequency ranges, where high-quality dielectrics are available.**

Transformation optics has been proposed as a general technique to design complex electromagnetic media with exotic properties, opening up paths for effectively rerouting light around an object to be concealed [1-3]. Experimental realization of a transformation-based invisibility cloak was first reported in microwave frequencies with metallic metamaterials [4]. Since singular material properties require resonant metamaterials, such cloaks only work in a very narrow frequency range and inevitably suffer from a strong absorption. A number of approaches are proposed to mitigate the material parameter constraints[5-16], among which the idea of carpet cloak[5] wins most attentions. Instead of transforming the cloaked area to a point invisible to our eyes, a carpet cloak disguises the obstacle from light by making it appear like a flat ground plane. The material parameters are nonsingular and can in principle be realized without the need of resonant features. Subsequent experimental realizations of this carpet cloak have been carried out in microwave [6,7] and optical frequencies [8-11] in both 2D [6, 8-10] and 3D [7, 11] scenarios.

In those implementations of carpet cloaks, a quasi-conformal mapping technique is taken which neglects the anisotropy of parameters, and structures with spatially varying densities are used to achieve the inhomogeneous isotropic parameter profiles. However, the spatially variant refractive index represents a complex design process resulting in long implementation cycles. In addition, theoretical work shows that the quasi-conformal mapping for eliminating the anisotropy will result in a lateral shift of the scattered wave, which is comparable to the height of the cloaked object, thereby making the object eventually detectable [17]. Furthermore, the quasi-conformal mapping limits the ratio of the size of the object to that of the cloak, since a large ratio corresponds to large anisotropy of the material parameters, breaking the prerequisite of this approximation. Here, we implement a different type of carpet cloak at optical frequencies employing a nano-structured artificial anisotropic material. The parameters of the cloak are homogeneous and rigorously designed [18], thus not compromising the performance of the cloaking. Such a homogeneous cloak has recently been introduced in macroscopic scale based on a natural birefringent crystal and experimentally demonstrated at visible frequencies [19, 20]. However, since the anisotropy of a natural material is usually low and cannot be engineered as desired, the hidden object has to be relatively small compared with the cloak device. In others words, macroscopic objects can only be concealed at the cost of using considerably larger space (orders of magnitude larger than the object itself). This challenge becomes more critical for applications in electronics industry, for example, in microcircuits and microsystems, wherein hiding the defects on intricate stencils or masks requires delicate and tiny devices. Our work, which is implemented on a silicon-on-insulator (SOI) platform, introduces a flexible way to address this concern. The high-anisotropic material composing the cloak is obtained by patterning the top silicon layer of an SOI wafer with nano-gratings of appropriate filling factor, which leads to a cloaked area only a few times smaller than the cloak itself. By precisely restoring the path of the reflecting wave from the surface, the cloak creates an illusion of a flat plane for a triangular bump on the surface, hiding its presence over a range of optical frequencies (measured between 1480nm to 1580nm).

The cloak takes a triangular shape as shown in Fig. 1 a. In a background with a permittivity $\varepsilon_b$, the permittivity tensor of a non-magnetic cloak ($\mu=1$) for transverse-magnetic (TM, magnetic field perpendicular to the cloak device) polarization can be expressed as [17]:

$$\bar{\bar{\varepsilon}} = \varepsilon_b \begin{bmatrix} \left(\frac{H_2}{H_2-H_1}\right)^2 & \pm\left(\frac{H_2}{H_2-H_1}\right)^2 \frac{H_1}{D} \\ \pm\left(\frac{H_2}{H_2-H_1}\right)^2 \frac{H_1}{D} & 1+\left(\frac{H_2}{H_2-H_1}\right)^2 \left(\frac{H_1}{D}\right)^2 \end{bmatrix}, \quad (1)$$

where $H_1$ and $H_2$ represent the heights of the obstacle and the cloak, respectively, and $D$ is half the bottom length of the cloak. The above problem can be diagonalized to $\begin{bmatrix} \varepsilon_\parallel & 0 \\ 0 & \varepsilon_\perp \end{bmatrix}$ by rotating the optical axis by the angle $\theta$. According to the effective medium theory, the anisotropic parameters can be realized with alternating layered materials[21], under the condition that the period of the alternating layers is much smaller than the wavelength. Therefore, appropriately combining two dielectric layers with permittivities $\varepsilon_1$ and $\varepsilon_2$ would yield the material with desired effective permittivity tensors: $\varepsilon_\parallel = r\varepsilon_1 + (1-r)\varepsilon_2$ and $\varepsilon_\perp = \dfrac{\varepsilon_1 \varepsilon_2}{r\varepsilon_2 + (1-r)\varepsilon_1}$, where $r$ denotes the filling factor of the constituent materials. The cloak parameters can be tailored by tuning the filling factor ($r$) and the orientation of the layers ($\theta$). As an example, Fig. 1 b shows the effective anisotropy $\varepsilon_\parallel / \varepsilon_\perp$ as a function of the filling factor $r$ for alternating air/silicon layers at $\lambda = 1500$ nm. At this frequency, the maximum anisotropy of 2.6 (corresponding to a birefringence $n_e - n_o = -0.847$) cannot be supported by any known, naturally occurring birefringent crystals. Fig. 1 b shows $H_1/D$ and $H_1/H_2$ in terms of the filling factor $r$, indicating that the shape of the cloak can be designed according to the requirement by appropriately arranging the composing materials. Therefore, layered materials bypass the limitation of natural material at hand and give us extra freedom to design the devices as desired.

In our experimental implementations, the carpet cloak (Fig. 1a) is realized by uniform silicon gratings fabricated with the electron-beam lithography (EBL) and reactive ion etching processes on an SOI wafer (250nm crystalline Si layer on top of a 3$\mu$m SiO$_2$ over a bulk Si substrate). The reflecting surface (the etched side-walls of the yellow region in Fig. 2 a) is coated with gold. In the vertical direction, the light confinement is achieved with index guiding, and the effective index (2.95) of the guided fundamental mode in the top silicon layer is used as $\varepsilon_1$ in the 2D designs. Fig. 2 b shows a scanning electron microscope image of the cloak device. The bottom width and the height of the cloak are 10$\mu$m and 8$\mu$m, respectively. The triangular concealed region where objects eventually can be hidden has a width of 10μm and a height of 1.84μm. A tapered input waveguide is used to guide the beam towards the cloak at a direction perpendicular to the lateral side of the cloak triangle. The end of the input waveguide has a width of the 6μm and a distance of 5μm from the cloak. The output beam is diffracted to the vertical direction by a photonic crystal grating coupler [22], and then eventually imaged by an infrared camera in order to observe its pattern. The distance of the grating from the cloak is 24μm. The insets show the oblique view of the carpet cloak (top) and the cloak/reflector interface (bottom). The period of the gratings composing the cloak is 140nm, and the filling factor *r* of Si is 0.5. In the measurement, the light with TM polarization from a tunable laser (1480nm to 1580nm) is used to characterize the cloak.

To illustrate how electromagnetic waves interact with the structure, we use a commercial finite-element method package (COMSOL Multiphysics) to simulate the propagation profile of a Gaussian beam reflected from a flat plane, a cloaked protruded plane, and a bare protruded plane. Fig 3 shows the magnetic field distributions for these three cases. The beam is split into two largely separated beams after reaching the protrusion of the ground surface (Fig. 3b), whereas in presence of the cloak constructed with grating structures, the reflection profile closely resembles that reflected by a flat surface (Fig. 3a).

The measurement of the cloak device is carried out for wavelengths ranging from 1480nm to 1580nm (restricted by the bandwidth of the tunable laser source), and Fig. 3 shows the result for 1500nm wavelength, at which the cloak is designed. The left panels show the images for the three configurations captured by the infrared camera, and the right panels show the corresponding intensity profiles at the output gratings. The profiles of the beam reflected by the flat mirror (Fig. 4a) and that reflected by the cloaked bump (Fig. 4b) are almost identical. In contrast, the reflection from an uncloaked bump (Fig. 4c) shows a lateral shift and a shape deformation. These experimental results serve as direct evidence that the cloak has effectively changed the path of the beam and given the observer an illusion of a flat surface. Consequently, any object hidden underneath the bump would be invisible for external observers. It should be noted that the cloak was designed for a background with a refractive index n=1.5 instead of silicon used in the experiment. Since we have deliberately relaxed impedance matching, reflection occurs at the interface between the cloak and the silicon background. However, since the direction of the incident beam is normal to the interface, the mismatch of impedance would not affect the direction of the input and output beams in the flat surface and cloak cases.

Experimental results for several other frequencies (1480nm, 1550nm, and 1580nm) are displayed in Fig. 5. The microscopic images clearly demonstrate that the effectiveness of the cloak is unaffected by varying frequencies. Although the cloak was measured from 1480nm to 1580nm, it in principle operates in a broad band, since silicon shows very modest dispersion over a wide frequency range. On the other hand, the waveguide dispersion due to the vertical index guiding is dominant in the present implementation. Nevertheless, the overall dispersion is still much less than that of a resonance-based metamaterials (From 1400nm to 1800nm, the effective index of silicon varies from 2.81 to 2.99), and the waveguide dispersion can be reduced or engineered by changing the thickness of the top silicon layer or using some over-cladding layer. Since the cloak is made exclusively of dielectric materials which are highly transparent to infrared light, and does not rely on the resonance of the metamaterial, the cloak itself absorbs a negligible fraction of energy. The reduced transmission observed in the measurement is mainly caused by the reflection at the interface and partly resulted by the imperfection of the fabrication. Consequently, adding an index matching layer around the cloak and improving uniformity of the grating would help eliminate the reflection and scattering.

In our experiment, the ratio of the size (area) of the bump to that of the cloak device is 0.23. Compared with the previous carpet cloaks in infrared frequencies[8-11], where the ratios are roughly smaller than 0.01, and the recently reported calcite cloaks by Birmingham[19] and MIT groups[20], which have ratios of 0.06 and 0.088, respectively, our cloak considerably economizes on the fabrication space. Moreover, in microwave frequencies where we have access to dielectric materials with very high permittivity, this ratio can be even more significantly increased. For instance, with a ferroelectric ceramic, Ba0:5Sr0:5TiO3, which has a relative dielectric constant of about 600, the effective anisotropy of the dielectric/air alternating layers ($\varepsilon_\parallel / \varepsilon_\perp$) can reach up to 150, which corresponds to an object/cloak ratio of 0.9.

We have designed and experimentally realized a homogeneous invisibility cloak at optical frequencies. In contrast to the previous works, the proposed device shows advantages of easier fabrication process and larger concealing area relative to the cloak device. Although this experiment is carried out in infrared frequencies, this design strategy is well applicable in other frequency ranges. Given proof of this concept, we would anticipate that with more precise fabrication our strategy should also yield a true invisibility carpet that works in the microwave and visible parts of the spectrum, and at a larger size, with the promise of a host of futuristic applications in industries and defence.

# References

1. Pendry, J. B., Schurig, D. & Smith, D. R. Controlling electromagnetic fields. Science, 312, 1780–1782 (2006).

2. Leonhardt, U., Optical conformal mapping. Science 312, 1777-1780 (2006).

3. Greenleaf, A., Lassas M. & Uhlmann G, Anisotropic conductivities that cannot be detected by EIT, Physiological Measurement, 24, 413 (2003).

4. Schurig, D. et al. Metamaterial electromagnetic cloak at microwave frequencies. Science 314, 977-980 (2006).

5. Li, J. & Pendry, J. B. Hiding under the carpet: A new strategy for cloaking. Phys. Rev. Lett. 101, 203901 (2008).

6. Liu, R. et al. Broadband ground-plane cloak. Science 323, 366-369 (2009).

7. Ma, H. F. & Cui, T. J., Three-dimensional broadband ground-plane cloak made of metamaterials, Nature Communications, 1, 21 (2010)

8. Valentine, J., Li, J., Zentgraf, T., Bartal, G. & Zhang, X., an optical cloak made of dielectrics, Nat. Mater. 8, 568 (2009).

9. Gabrielli, L. H., Cardenas, J., Poitras, C. B. & Lipson, M., Silicon nanostructure cloak operating at optical frequencies, Nat. Photonics 3, 461 (2009).

10. Lee, J. H., Blair, J., Tamma, V. A., Wu, Q., Rhee, S. J., Summers, C. J.,and Park, W., Direct visualization of optical frequency invisibility cloak based on silicon nanorod array, Opt. Express 17, 12922-12928, 2009.

11. Ergin, T., Stenger, N., Brenner, P., Pendry, J. B. & Wegener, M., Three-dimensional invisibility cloak at optical wavelengths, Science, 328, 337 (2010).

12. Cai, W., Chettiar, U. K., Kildishev, A. V. & Shalaev, V. M. Optical cloaking with metamaterials. Nature Photon. 1, 224_227 (2007).

13. Tretyakov, S., Alitalo, P., Luukkonen, O. & Simovski, C., Broadband electromagnetic cloaking of long cylindrical objects, Phys. Rev. Lett., 103, 103905 (2009)

14. Smolyaninov, I. I., Smolyaninova, V. N., Kildishev, A. V. & Shalaev, V. M., Anisotropic metamaterials emulated by tapered waveguides: Application to optical cloaking, Phys. Rev. Lett. 102, 213901 (2009).

15. Leonhardt, U. & Tyc, T. Broadband invisibility by non-euclidean cloaking. Science, 323, 110 (2009).

16. Alu, A.&Engheta. N, Multifrequency optical invisiblitity cloak with layered plasmonic shell. Phys. Rev. Lett. 100, 113901, (2008).

17. Zhang, B. L., Chan, T. & Wu, B. I., Lateral shift makes a ground-plane cloak detectable, Phys. Rev. Lett. 104, 233903 (2010).

18. Luo, Y., Zhang, J. J., Chen, H. S., Ran, L. X., Wu B. I. & Kong, J. A., A rigorous analysis of plane-transformed invisibility cloaks, IEEE transactions on antennas and propagation, 57, 3926 (2009)

19. Chen, X. Luo, Y. Zhang, J. J., Jiang, K. Pendry, J. B. and Zhang S., Macroscopic invisibility cloaking of visible light, arXiv:1012.2783.

20. Zhang, B. Luo, Y. Liu, X. Barbastathis, G. Macroscopic Invisible Cloak for Visible Light, Phys. Rev. Lett. 106, 033901 (2011).

21. Wood, B. Pendry, J. B. and Tsai, D. P. Directed subwavelength imaging using a layered metal-dielectric system, Phys. Rev. B 74, 115116, 2006.

22. L, Liu. Pu, M. Yvind, K. & Hvam, J. M., High-Efficiency, Large-Bandwidth Silicon-on-Insulator Grating Coupler based on a Fully-Etched Photonic Crystal Structure, Appl. Phys. Lett. 96, 051126 (2010).

Fig. 1

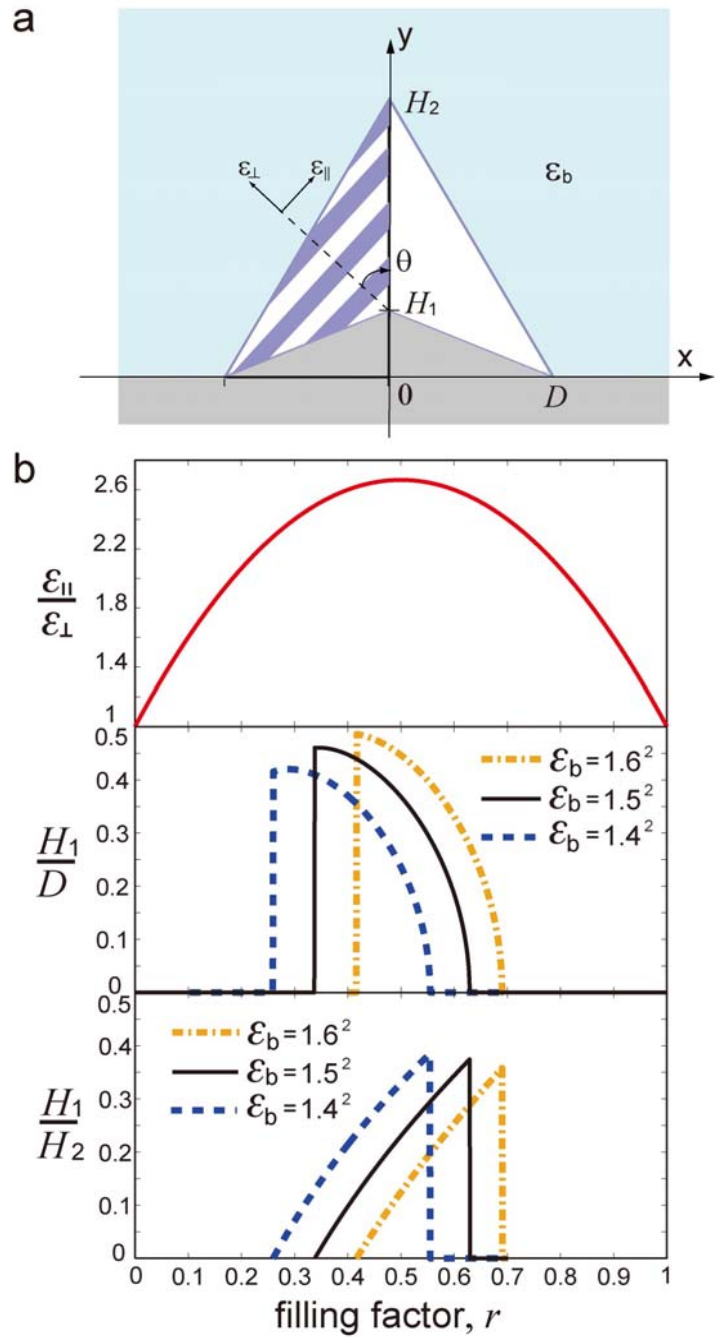

Fig. 1 (a) schematic of the proposed homogeneous cloak achieved with uniform layered medium. Here $\varepsilon_\parallel$ and $\varepsilon_\perp$ represent the two principle permittivity components of the effective anisotropic media. (b) the material anisotropy $\varepsilon_\parallel / \varepsilon_\perp$ (top panel), the relative height of the obstacle $H_1 / D$ (middle panel), and the obstacle/cloak ratio in terms of the height $H_1 / H_2$ (bottom panel) as functions of the filling factor of the grating $r$.

Fig. 2

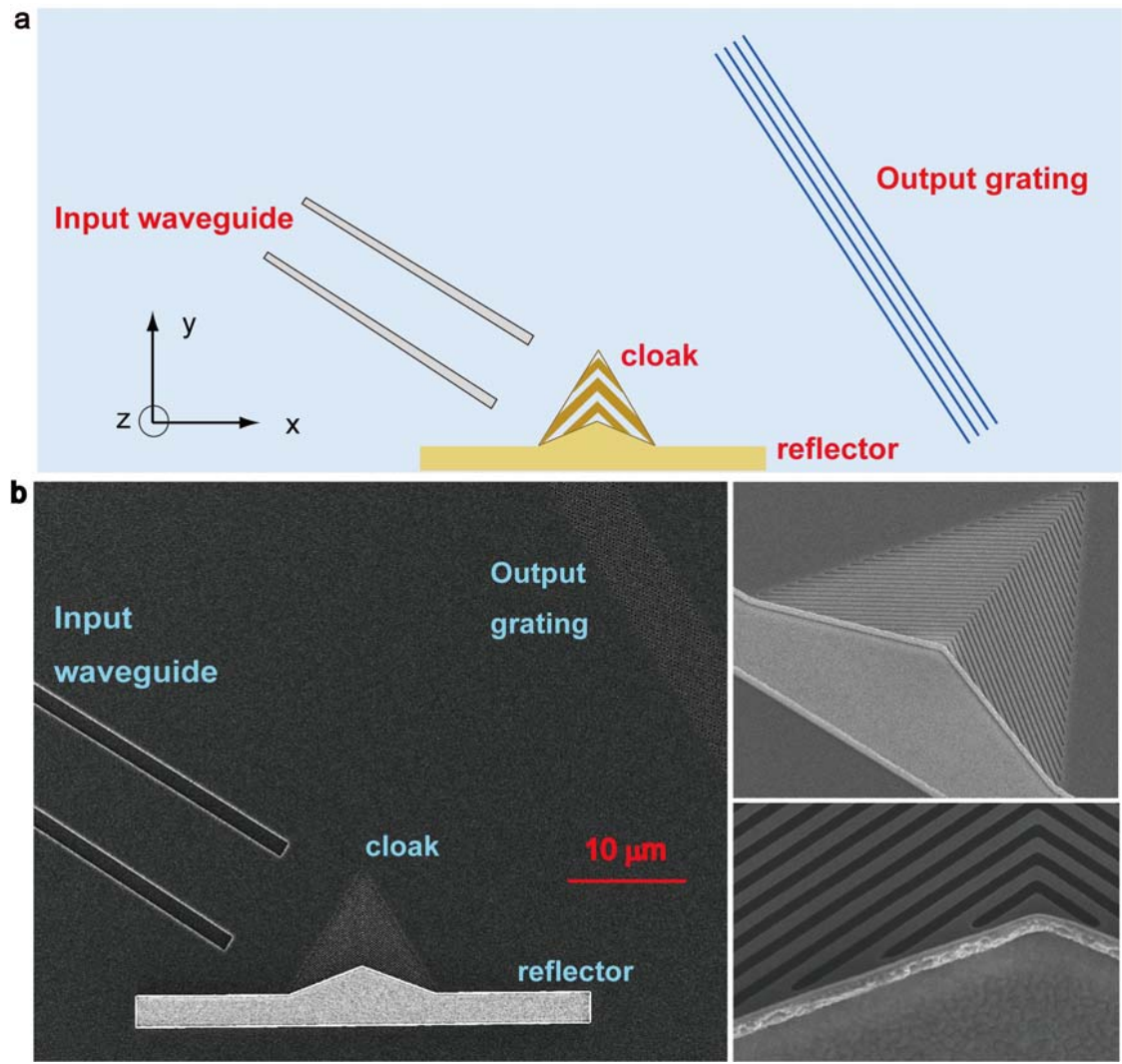

Fig. 2 The fabricated carpet cloak. (a). Schematic diagram of a fabricated carpet cloak. Light is coupled to the cloak through the input waveguide and reflected at the gold mirror. The reflected beam is detected by the output grating. (b). Scanning electron microscopic image of a fabricated carpet cloak, The insets show the oblique view of the carpet cloak (top) and the cloak/reflector interface (bottom).

Fig. 3

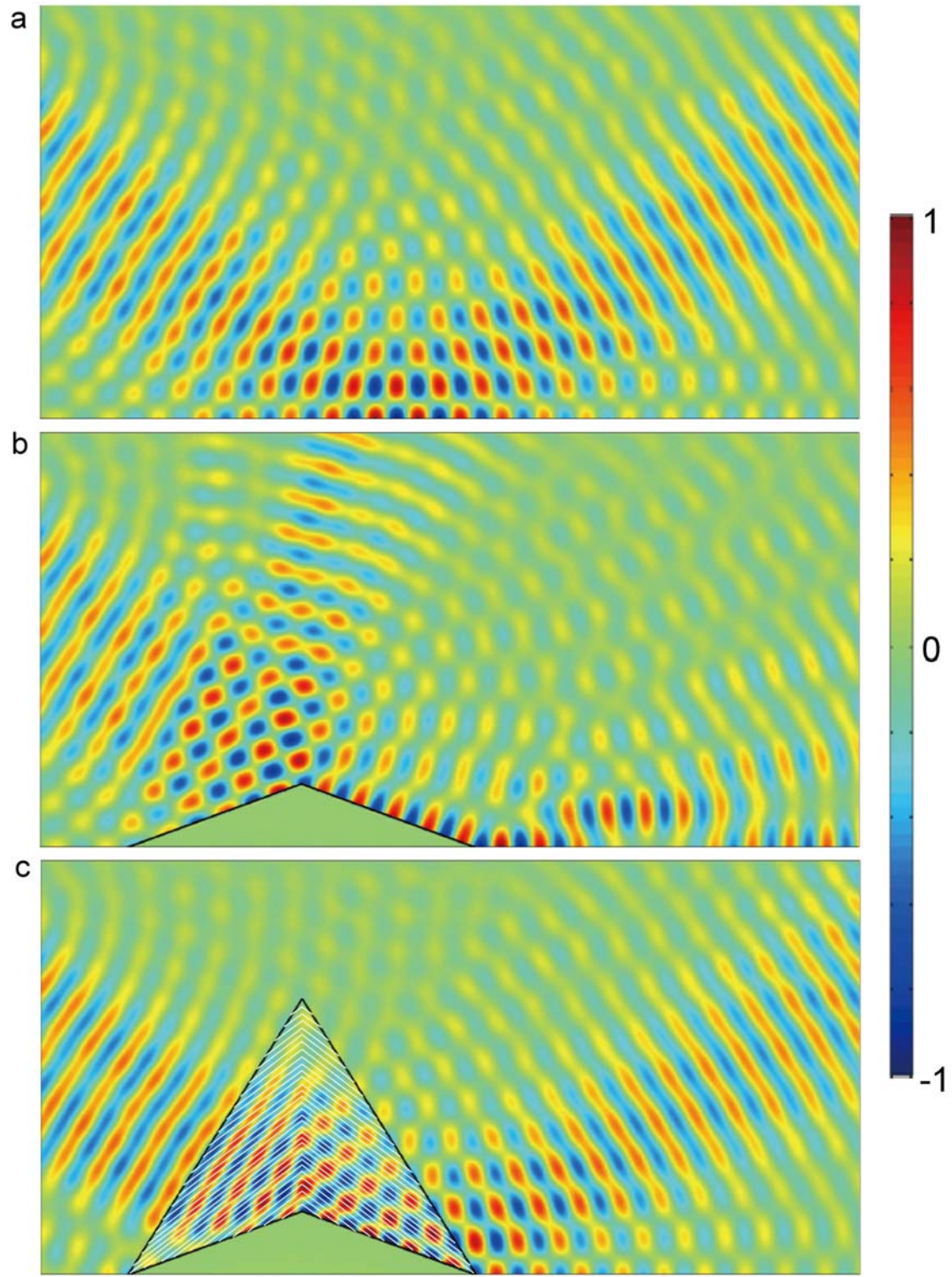

Fig. 3 Numerical simulations of the carpet cloak system at 1550nm. The refractive index of the background is 1.5. (a). A Gaussian beam is obliquely incident upon a reflecting surface. (b). A Gaussian beam is obliquely incident upon a reflecting surface with a cloaked triangular protrusion. The direction of the beam is perpendicular to a lateral surface of the cloak (c). A Gaussian beam is obliquely incident upon a reflecting surface with a triangular protrusion.

Fig. 4

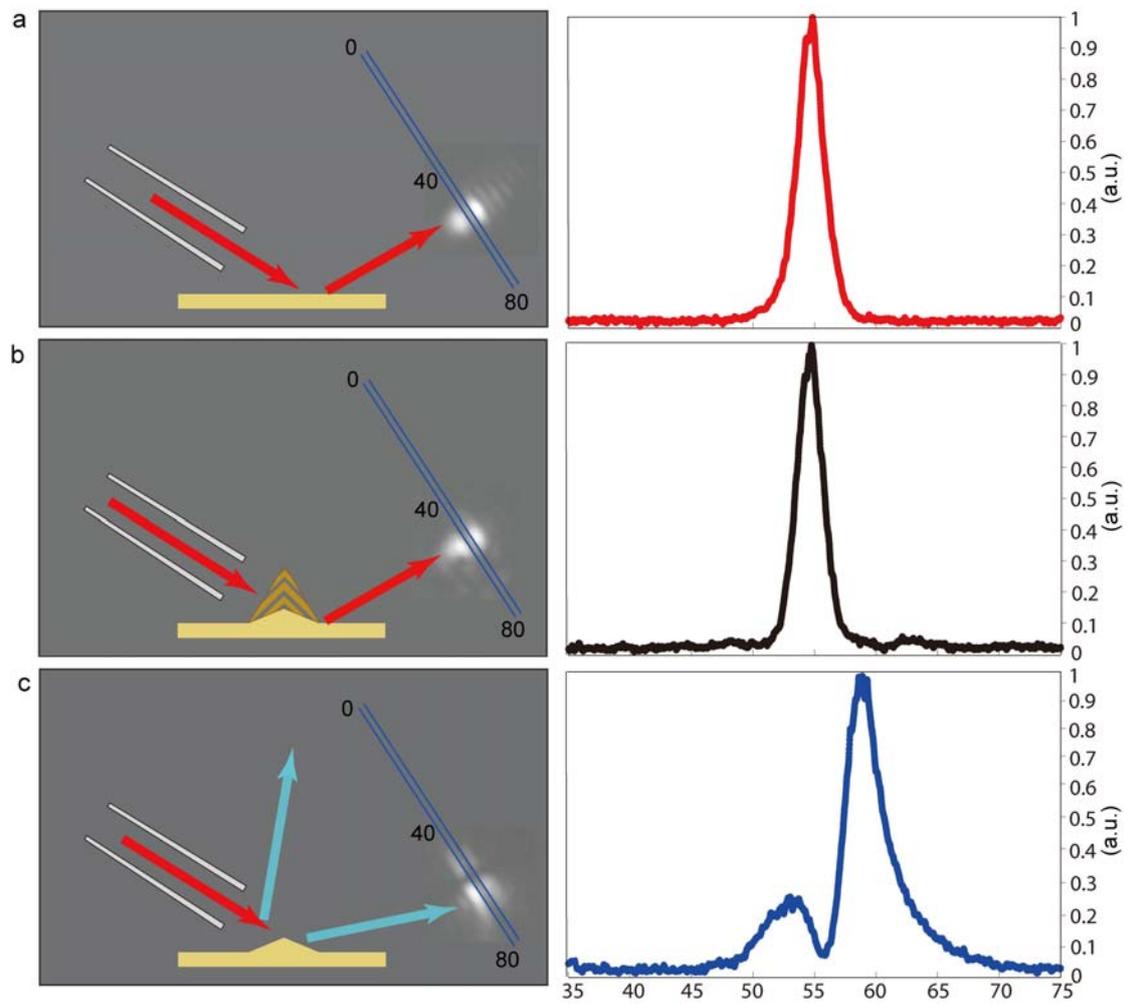

Fig. 4 The measured output image from the fabricated devices (left) and the plot of intensity along the output grating (right) at 1500nm. The results for a Gaussian beam reflected from a flat surface(a), a protruded surface (b), and the same protruded surface with a cloak.

Fig. 5

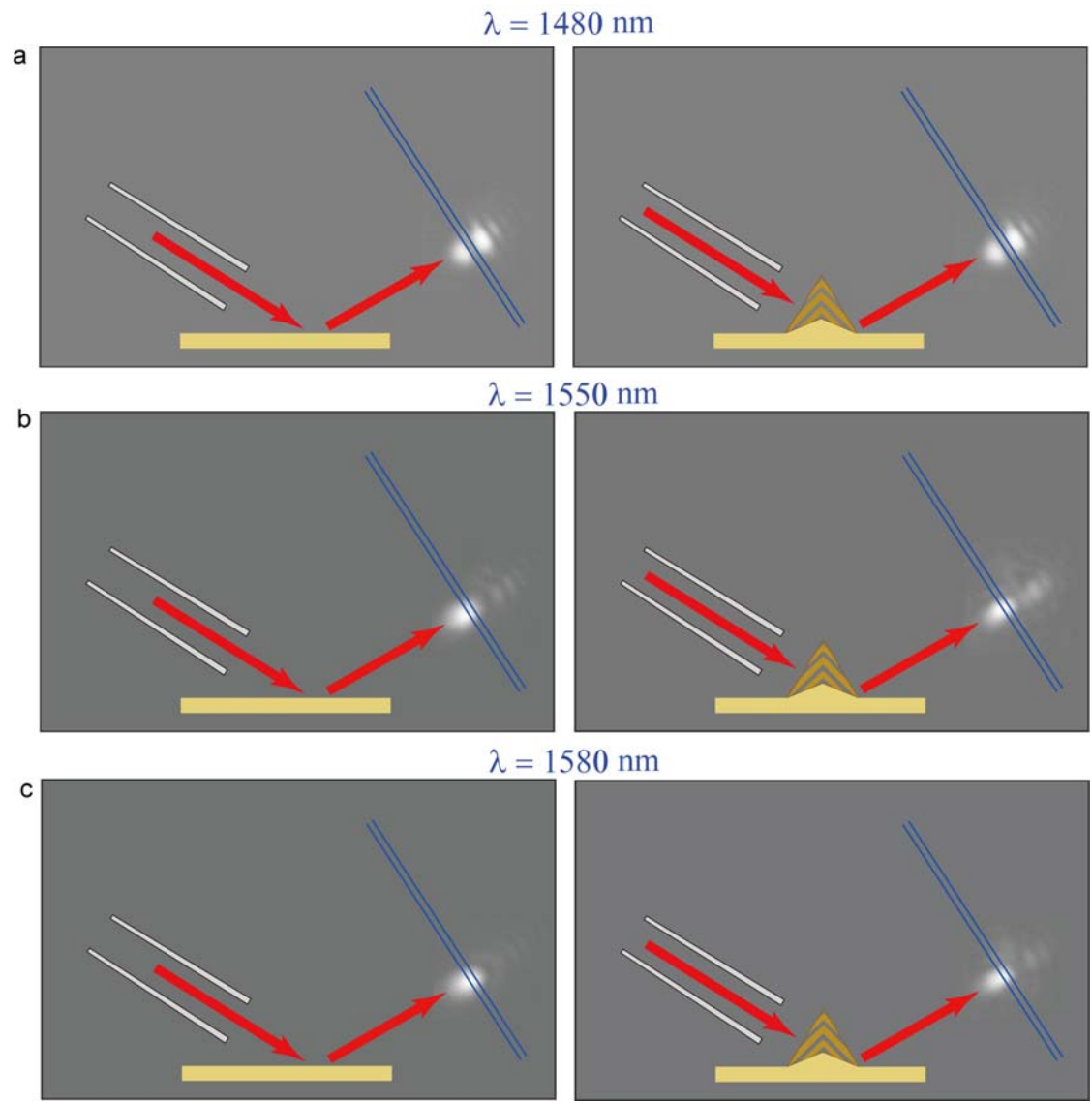

Fig. 5 The measured output image from a flat surface (left) and a cloaked protruded surface (right) at 1480nm (a), 1550nm (b), and 1580nm (c).